\documentclass[11pt,sort&compress]{elsarticle}
\usepackage{amsmath,amssymb,latexsym,epsfig,multirow,graphicx,longtable,rotating, hyperref, xcolor}

\begin{document}

\title{Temporal and Spacial Studies  of Infectious Diseases: \\
Mathematical Models and Numerical Solvers}

\author[google]{Yongjia~Xu}
\ead{bellaxu225@gmail.com}

\author[smu]{Md Abu~Talha}
\ead{atalha@smu.edu}

\author[ua]{Shan~Zhao}
\ead{szhao@ua.edu}

\author[smu]{Weihua~Geng\corref{cor1}}
\ead{wgeng@smu.edu}

\cortext[cor1]{Corresponding author}

\address[google]{Google LLC, Bella Xu / bellaxu@, 1600 Amphitheater Pkwy, Mountain View, CA 94043}
\address[smu]{Department of Mathematics, Southern Methodist University, Dallas, TX 75275 USA}
\address[ua]{Department of Mathematics, University of Alabama, Tuscaloosa, AL 35487 USA}

\begin{abstract}
\indent 
The SIR model is a classical model characterizing the spreading of infectious diseases. 
This model describes the time-dependent quantity changes among Susceptible, Infectious, and Recovered groups. 
By introducing space-depend effects such as diffusion and creation in addition to the SIR model, the Fisher's model is in fact a more advanced and comprehensive model. 
However, the Fisher's model is much less popular than the SIR model in simulating infectious disease numerically
due to the difficulties from 
the parameter selection,  
the involvement of 2-d/3-d spacial effects, 
the configuration of the boundary conditions, etc.

This paper aim to address these issues 
by providing numerical algorithms involving space and time finite difference schemes and iterative methods, 
and its open-source Python code for solving the Fisher's model. 
This 2-D Fisher's solver is second order in space and up to the second order in time, 
which is rigorously verified using test cases with analytical solutions.
Numerical algorithms such as SOR, implicit Euler, Staggered Crank-Nicolson, and ADI are combined to improve the efficiency and accuracy of the solver. 
It can handle various boundary conditions subject to different physical descriptions. 
In addition, real-world data of Covid-19 are used by the model to demonstrate its practical usage 
in providing prediction and inferences. 

\vspace*{1cm}


\end{abstract}
\date{\today}
\maketitle

\section{Introduction}



\noindent
\textcolor{blue}{
- Read the first a few sections of the two chapters in Murray's model to write an introduction to SIR and Fisher's model (wording, no formula). \\
- Some related paper: 1-d model \cite{Kenkre:2024}; 2-d model with analytical solution \cite{Lou:2023}, 2-d lattice method \cite{Vaziry:2022}, 2-d cell-sheet wound closure \cite{Habbal:2014}, extended 1-d and 2-d model \cite{Li:2009}. Try to find more !!!\\
When reading and searching references: focusing on two problems.\\
- What kind of boundary value conditions are used? \\
- How did the practical data are used with the mathematical model?\\
}

In this paper, we solve the 2-D Fisher's model using centered difference for spacial discretization and 
implicit Euler method and staggered Crank-Nicolson (CN) method, which maintains the 2nd order convergence in space and first or second order convergence in time. 
In each time step of CN method, 
the spatially discretized system is solved using 
full matrix with direct method, 
SOR method, 
and ADI method. 

The SOR method solves the problem in $O(n^3)$, 
where $n$ is the number of unknowns in each direction. 
The advantages of SOR method are that 
we do not need to explicitly build the linear algebraic matrix $A$, 
which is complicated in algorithm and expensive in memory usage. 
Furthermore, when different boundary conditions are involved, 
SOR methods only requires small modification. 

The ADI method solves the problem in $O(n^2)$. ADI method introduce some truncation error but avoids the iteration errors. 

\noindent
Our reasearch focus are:\\
(1) How are the boundary conditions are used and treated in numerical framework?\\
(2) How are practical data involved with the mathematical model. \\

We provide the python code for solving the Fisher's model in 2-D with three typical boundary conditions, which will provide a powerful tool for bio-scientists and healthcare researchers to simulate infectious diseases. 

We first validate the solvers using designed cases with possibly available analytical solutions.

\section{Theory and Algorithms}
\label{theory}

\subsection{Theory}
The most well-recognized mathematical model in simulating infectious diseases are the time-dependent SI model (or SIR and SEIR if recovery and incubation are considered) and the time and space-dependent Fisher's model.

\subsubsection{1-D SIR model}
The 1-D SIR model is given as:\\
\begin{equation}
\displaystyle \left\{\begin{array}{ll} 
\displaystyle\frac{dS}{ d t} = & -rSI \\ 
\displaystyle\frac{d I}{ dl t} = & rSI-aI \\
\displaystyle\frac{dR}{dt}=&aI
\end{array}\right. \label{eq_SIR}
\end{equation}
with $S(t)$, $I(t)$, and $R(t)$ as the number of individuals in the susceptibles, the infectives, and the recovered. The parameter $r>0$ is the infection rate and $a>0$ the removal rate of infectives. Since this model is well known, studied, and used by educator, researcher, and practitioner, we will only briefly emphasize a few aspects related to the model. \\
First, since the rate of change of $R$ is only related to $I$, for simplicity, we may only consider the $\displaystyle\frac{dS}{dt}$ and the $\displaystyle\frac{dI}{dt}$ terms in Eq.~(\ref{eq_SIR}) with the $-aI$ term removed. Such a further simplified model is known  as the $SI$ model. Under the assumption that $S+I=N$ the total population, $S(t)$ can then be reduced as
\begin{equation}
    \displaystyle\frac{dS}{dt}=-rS(N-S).
\end{equation}
Second, the solution to this equation is 
 analytically available as the logistic function 
 \begin{equation}
 S(t)=\frac{N}{1+\frac{1}{C}e^{-Nrt}}
 \end{equation}
 which can be derived using the change of variables strategy in solving Bernoulli's equation or using the partial fractions. Note $I(t) = N - S(t)$ will have similar but reversed pattern. If $R(t)$ is considered, we can easily figure out  
$\displaystyle R_0=\frac{r}{a}S_0$, which is known as the basic reproduction rate, the number of secondary infections produced by one primary infection in a wholly susceptible population.

\subsubsection{2-D  Fisher's model}
Here we briefly introduce the 2-D Fisher's model \cite{Fisher:1937}. 

\begin{equation}
\displaystyle \left\{\begin{array}{ll} \displaystyle\frac{\partial S}{ \partial t} = & -k_SIS+\displaystyle d_S\left(\frac{\partial^2S}{\partial x^2} +\displaystyle \frac{\partial^2S}{\partial y^2}\right) + f(x,y,t) \\
\\
\displaystyle\frac{\partial I}{ \partial t} = & - k_II(\lambda-S)  + \displaystyle d_I\left(\frac{\partial^2I}{\partial x^2}+\displaystyle \frac{\partial^2I}{\partial y^2}\right) + g(x,y,t) \\
\end{array}\right. \label{eq_2dFisher}
\end{equation}
In this equation $S(t,x,y)$ and $I(t,x,y)$ are the population (or population density if normalized) of the susceptible group (S) and infectious group (I) in a region, $t$ is the elapsed time, $x, y$ are the 2-d spacial coordinates, and $k_S,k_I,d_S,d_I$ are reaction rate constants and diffusion constants for $I$ and $S$. The function $f$ and $g$ can enforce the increment or decrement of $S$ and $I$ if needed. 
Additionally, initial and boundary conditions are added 
according to the practical data. Rate constants can also be included as parameters, calibrated using real data. 

\subsection{Discretization with finite difference methods}
We will solve the Fisher's equation numerically. 
Time discretization will be treated with 
Explicit Euler (simple, conditionally stable, first order), 
Implicit Euler (complicated, unconditionally stable, first order),  
and Crank-Nicolson (complicated, unconditionally stable, second order) 
as options with trade off of cost, efficiency, and accuracy. 
Space is discretized using centered difference. 
Three type of boundary conditions: 
Dirichlet (fixed values) , 
Riemann (fixed rate of change), 
and Robin (a mix of both) 
will be applied to model the disease spreading across regions. 

In each time step, the space-discretized linear algebraic system will be solved using the SOR method. For a 2-D problem from the Fisher's model, for each time step the computational cost using SOR with optimized $\omega$ is $O(N^3)$ as opposed to the $O(N^6)$ for Gaussian Elimination, where $N$ is the number of unknowns in each direction.

\noindent
For the simplification, we slightly modified the 2-d fisher's equation as shown in (\ref{eq_2dFisher}) by setting all coefficients as 1. 
The resulting differential equation is given as:\\
\begin{equation}
\displaystyle \left\{\begin{array}{ll} \displaystyle\frac{\partial S}{ \partial t} = & -IS+\displaystyle \frac{\partial^2S}{\partial x^2} +\displaystyle \frac{\partial^2S}{\partial y^2} + f(x,y,t) \\
\\
\displaystyle\frac{\partial I}{ \partial t} = & - I(\lambda-S)  + \displaystyle \frac{\partial^2I}{\partial x^2}+\displaystyle \frac{\partial^2I}{\partial y^2} + g(x,y,t) \\
\end{array}\right. \label{eq_Fisher2d}
\end{equation}

This equation needs to be discretized in both space and time to receive numerical solution. We spatially discretize it using centered finite difference scheme for a second order convergence. For time discretization we use implicit Euler scheme for a first order convergence and staggered Crank-Nicolson scheme for a second convergence. The implicit Euler scheme is simple and direct while the staggered CN scheme requires more complicated treatment. Since there are two coupled variables $S$ and $I$, the staggered scheme as detailed below becomes necessary to secure the second order convergence.

\subsubsection{Implicit Euler method}

For $i,j=1,\cdots,n$ and $k=1,\cdots, m$, at the $(k+1)$th time step, we have the discretized Fisher's equation using centered difference in space and Implicit Euler in time as\\
\begin{equation}
\displaystyle \left\{\begin{array}{ll} \displaystyle\frac{S_{ij}^{k+1}-S_{ij}^k}{ \Delta t} 
&= -I_{ij}^k S_{ij}^{k+1}+\displaystyle \frac{
-4S^{k+1}_{ij}+S^{k+1}_{i-1,j}+S^{k+1}_{i+1,j}+S^{k+1}_{i,j-1}+S^{k+1}_{i,j+1}}{h^2} + f_{ij}^{k+1} \\
\\
\displaystyle\frac{I_{ij}^{k+1}-I_{ij}^k}{ \Delta t} &= - I_{ij}^{k+1}(\lambda-S_{ij}^{k}) + \displaystyle \frac{-4I^{k+1}_{ij}+I^{k+1}_{i-1,j}+I^{k+1}_{i+1,j}+I^{k+1}_{i,j-1}+I^{k+1}_{i,j+1}}{h^2} + g_{ij}^{k+1} \\
\end{array}\right.
\end{equation}

\noindent Let $l = \frac{\Delta t}{(\Delta x)^2}$ and rearrange by time step as 
\begin{equation}
\displaystyle \left\{\begin{array}{ll} 
-lS^{k+1}_{i-1,j}-lS^{k+1}_{i,j-1}+(1+4l+I_{ij}^k \Delta t)S^{k+1}_{ij}-lS^{k+1}_{i+1,j}-lS^{k+1}_{i,j+1}
&= S_{ij}^k + f_{ij}^{k+1} \Delta t \\
\\

-lI^{k+1}_{i-1,j}-lI^{k+1}_{i,j-1}+(1+4l+(\lambda-S_{ij}^k) \Delta t)I^{k+1}_{ij}-lI^{k+1}_{i+1,j}-lI^{k+1}_{i,j+1}
&= I_{ij}^k + g_{ij}^{k+1}\Delta t \\
\end{array}\right.
\end{equation}

\subsubsection{Staggered Crank-Nicolson method}

\noindent
For $i,j=1,\cdots,n$ and $k=1,\cdots, m$, the $k+1$th time-step at the grid $(x_i,y_j)$, the discretized form using staggered CN for time and centered difference for space is given as:\\
\begin{equation}
\displaystyle \left\{\begin{array}{ll} \displaystyle\frac{S_{ij}^{k+1}-S_{ij}^k}{ \Delta t} 
&= -I_{ij}^{k+\frac{1}{2}} \displaystyle \frac{S_{ij}^{k+1}+S_{ij}^{k}}{2}+\displaystyle \frac{
-4S^{k}_{ij}+S^{k}_{i-1,j}+S^{k}_{i+1,j}+S^{k}_{i,j-1}+S^{k}_{i,j+1}}{2h^2}\\
&+\displaystyle \frac{
-4S^{k+1}_{ij}+S^{k+1}_{i-1,j}+S^{k+1}_{i+1,j}+S^{k+1}_{i,j-1}+S^{k+1}_{i,j+1}}{2h^2} + \displaystyle\frac{f_{ij}^{k+1}+f_{ij}^{k}}{2} \\

\displaystyle\frac{I_{ij}^{k+\frac{1}{2}}-I_{ij}^{k-\frac{1}{2}}}{ \Delta t} &= - \displaystyle\frac{I_{ij}^{k+\frac{1}{2}}+I_{ij}^{k-\frac{1}{2}}}{2}(\lambda-S_{ij}^{k}) + \displaystyle \frac{-4I^{k-\frac{1}{2}}_{ij}+I^{k-\frac{1}{2}}_{i-1,j}+I^{k-\frac{1}{2}}_{i+1,j}+I^{k-\frac{1}{2}}_{i,j-1}+I^{k-\frac{1}{2}}_{i,j+1}}{2h^2}\\
&+ \displaystyle \frac{-4I^{k+\frac{1}{2}}_{ij}+I^{k+\frac{1}{2}}_{i-1,j}+I^{k+\frac{1}{2}}_{i+1,j}+I^{k+\frac{1}{2}}_{i,j-1}+I^{k+\frac{1}{2}}_{i,j+1}}{2h^2} + \displaystyle\frac{g_{ij}^{k+\frac{1}{2}}+g_{ij}^{k-\frac{1}{2}}}{2} \\
\end{array}\right. \label{eq_CN1}
\end{equation} 
Note here $S$ is evaluated at $k$ and $k+1$ in time while $I$ is evaluated at $k-\frac{1}{2}$ and $k+\frac{1}{2}$ makes the scheme {\it staggered}, which avoids to evaluate $I_{ij}$ at both $k$ and $k+1$ for the first equation and evaluate $S_{ij}$  at both $k$ and $k+1$ for the second equation for the regular C-N scheme (i.e. the trapezoid rule to evaluate the integrand function at $t_k$ and $t_{k+1}$).

\noindent Let $l = \frac{\Delta t}{2h^2}$ and rearrange by time step as \\
\begin{equation}
\displaystyle \left\{\begin{array}{ll} 
-lS^{k+1}_{i-1,j}-lS^{k+1}_{i,j-1}+(1+4l+\displaystyle\frac{I_{ij}^{k+\frac{1}{2}} \Delta t}{2})S^{k+1}_{ij}-lS^{k+1}_{i+1,j}-lS^{k+1}_{i,j+1} =
\\ -\displaystyle\frac{I_{ij}^{k+\frac{1}{2}} S_{ij}^k}{2} \Delta t+S_{ij}^k -4lS^{k}_{ij} + lS^{k}_{i-1,j} + lS^{k}_{i,j-1} 
+ lS^{k}_{i+1,j} + lS^{k}_{i,j+1} + \displaystyle\frac{f_{ij}^{k+1}+f_{ij}^{k}}{2}  \Delta t \\
-lI^{k+\frac{1}{2}}_{i-1,j}-lI^{k+\frac{1}{2}}_{i,j-1}+(1+4l+\displaystyle\frac{(\lambda-S_{ij}^k) \Delta t}{2})I^{k+\frac{1}{2}}_{ij}-lI^{k+\frac{1}{2}}_{i+1,j}-lI^{k+\frac{1}{2}}_{i,j+1} = \\
-\displaystyle\frac{(\lambda-S_{ij}^k) I_{ij}^{k-\frac{1}{2}} \Delta t }{2} + I_{ij}^{k-\frac{1}{2}} -4lI^{k-\frac{1}{2}}_{ij}+ lI^{k-\frac{1}{2}}_{i-1,j} + lI^{k-\frac{1}{2}}_{i,j-1} + lI^{k-\frac{1}{2}}_{i+1,j} + lI^{k-\frac{1}{2}}_{i,j+1} + \displaystyle\frac{g_{ij}^{k+\frac{1}{2}}+g_{ij}^{k-\frac{1}{2}}}{2}\Delta t
\end{array}\right. \label{eq_DisFisherCN}
\end{equation}
Here we will apply the basic iterative methods such as Jacobi, Gauss-Siedel, and SOR method to solve Eq.~(\ref{eq_DisFisherCN}), which has another layer of iterations in each time step. We need to simply the formulation to illustrate the ideas.

\noindent
Let 
$u_{ij}=S^{k+1}_{ij}$, $v_{ij}=I^{k+1/2}_{ij}$, 
$l_c=1+4l+\displaystyle\frac{I_{ij}^{k+\frac{1}{2}} \Delta t}{2}$, $l_d=1+4l+\displaystyle\frac{(\lambda-S_{ij}^k) \Delta t}{2}$
\\
$c_{ij}=-\displaystyle\frac{I_{ij}^{k+\frac{1}{2}} S_{ij}^k}{2} \Delta t+S_{ij}^k -4lS^{k}_{ij}+ lS^{k}_{i-1,j} + lS^{k}_{i,j-1}
+ lS^{k}_{i+1,j} + lS^{k}_{i,j+1}+ \displaystyle\frac{f_{ij}^{k+1}+f_{ij}^{k}}{2}  \Delta t,$\\
$d_{ij}=-\displaystyle\frac{(\lambda-S_{ij}^k) I_{ij}^{k-\frac{1}{2}} \Delta t }{2} + I_{ij}^{k-\frac{1}{2}}-4lI^{k-\frac{1}{2}}_{ij}+ lI^{k-\frac{1}{2}}_{i-1,j} + lI^{k-\frac{1}{2}}_{i,j-1} 
+ lI^{k-\frac{1}{2}}_{i+1,j} + lI^{k-\frac{1}{2}}_{i,j+1} + \displaystyle\frac{g_{ij}^{k+\frac{1}{2}}+g_{ij}^{k-\frac{1}{2}}}{2}\Delta t$\\

\noindent
The system to be solved in one time-step update is in this simplified form

\begin{equation}
\displaystyle \left\{
\begin{array}{ll} 
-lu_{i-1,j}-lu_{i,j-1}+l_c u_{ij}-lu_{i+1,j}-lu_{i,j+1}
&=c_{ij}\\
-lv_{i-1,j}-lv_{i,j-1}+l_d v_{ij}-lv_{i+1,j}-lv_{i,j+1}
&=d_{ij},
\end{array}\right.
\end{equation}
which can be solved by iterative methods such as Jacobi, G-S, and SOR as detailed below. For these iterative methods, to update the value of $u$ and $v$ at the $n$th iteration, we have: \\
\noindent 
Jacobi:\\
\begin{equation}
\displaystyle \left\{
\begin{array}{ll} 
u_{ij}^{(n+1)}= \frac{l}{l_c}\left((u_{i+1,j}^{(n)} + u_{i-1,j}^{(n)} + u_{i,j+1}^{(n)} + u_{i,j-1}^{(n)}) + c_{ij}/l\right)  \\
v_{ij}^{(n+1)}= \frac{l}{l_d}\left((v_{i+1,j}^{(n)} + v_{i-1,j}^{(n)} + v_{i,j+1}^{(n)} + v_{i,j-1}^{(n)}) + d_{ij}/l\right) 
\end{array}\right.
\end{equation}
G-S:\\
\begin{equation}
\displaystyle \left\{
\begin{array}{ll} 
u_{ij}^{(n+1)}= \frac{l}{l_c}\left((u_{i+1,j}^{(n)} + u_{i-1,j}^{(n+1)} + u_{i,j+1}^{(n)} + u_{i,j-1}^{(n+1)}) + b_{ij}/l\right)  \\
v_{ij}^{(n+1)}= \frac{l}{l_d}\left((v_{i+1,j}^{(n)} + v_{i-1,j}^{(n+1)} + v_{i,j+1}^{(n)} + v_{i,j-1}^{(n+1)}) + d_{ij}/l\right)
\end{array}\right.
\end{equation}SOR:\\
\begin{equation}
\displaystyle \left\{
\begin{array}{ll} 
l_cu_{ij}^{(n+1)}= l_cu_{ij}^{(n)} + \omega^*\left(-l_cu_{ij}^{(n)} + l(u_{i+1,j}^{(n)} + u_{i-1,j}^{(n+1)} + u_{i,j+1}^{(n)} + u_{i,j-1}^{(n+1)}) + c_{ij}\right)\\
l_dv_{ij}^{(n+1)}= l_dv_{ij}^{(n)} + \omega^*\left(-l_dv_{ij}^{(n)} + l(v_{i+1,j}^{(n)} + v_{i-1,j}^{(n+1)} + v_{i,j+1}^{(n)} + v_{i,j-1}^{(n+1)}) + d_{ij}\right)
\label{eqSOR1}
\end{array}\right.
\end{equation}
thus
\begin{equation}
\displaystyle \left\{
\begin{array}{ll} 
u_{ij}^{(n+1)}= u_{ij}^{(n)} + \omega^*\left(-u_{ij}^{(n)} + (l/l_c)(u_{i+1,j}^{(n)} + u_{i-1,j}^{(n+1)} + u_{i,j+1}^{(n)} + u_{i,j-1}^{(n+1)}) + c_{ij}/l_c\right)\\
v_{ij}^{(n+1)}= v_{ij}^{(n)} + \omega^*\left(-v_{ij}^{(n)} + (l/l_d(v_{i+1,j}^{(n)} + v_{i-1,j}^{(n+1)} + v_{i,j+1}^{(n)} + v_{i,j-1}^{(k+1)}) + d_{ij}/l_c\right)
\label{eqSOR2}
\end{array}\right.
\end{equation}

\subsubsection{ Alternating-Direction Implicit (ADI) Methods}
As mentioned earlier, for each time step, the SOR method can reduce the computational cost to $O(N^3)$ when the optimized $\omega$ is used. The Alternating-Direction Implicit (ADI) Methods can further improve the efficiency to $O(N^2)$ owing to the  Thomas's Algorithm to solve tridiagonal matrix. Below is the derivation. 

We rewrite Eq.~(\ref{eq_CN1}) as 
\begin{equation}
\displaystyle \left\{\begin{array}{ll} \displaystyle\frac{S_{ij}^{k+1}-S_{ij}^k}{ \Delta t} 
&= -I_{ij}^{k+\frac{1}{2}} \displaystyle \frac{S_{ij}^{k+1}+S_{ij}^{k}}{2} + (D_{xx}+D_{yy})\frac{S_{ij}^{k+1}+S_{ij}^k}{ 2}
+ \displaystyle{f_{ij}^{k+\frac{1}{2}}} \\

\displaystyle\frac{I_{ij}^{k+\frac{1}{2}}-I_{ij}^{k-\frac{1}{2}}}{ \Delta t} &= - \displaystyle\frac{I_{ij}^{k+\frac{1}{2}}+I_{ij}^{k-\frac{1}{2}}}{2}(\lambda-S_{ij}^{k}) + (D_{xx}+D_{yy})\displaystyle\frac{I_{ij}^{k+\frac{1}{2}}+I_{ij}^{k-\frac{1}{2}}}{2} + \displaystyle{g_{ij}^{k}}\\
\end{array}\right. \label{eq_CN-ADI1}
\end{equation}
where $D_{xx}$ and $D_{yy}$ are the central difference operators in $x$ and $y$ direction. For example, 
\begin{equation}
D_{xx} S_{ij}^k = \displaystyle \frac{ -2S^{k}_{ij}+S^{k}_{i-1,j}+S^{k}_{i+1,j}}{h^2}
\end{equation}
Using $f_{ij}^{k+\frac{1}{2}}$ is a simplified alternative to using $\frac{f_{ij}^{k+1}+f_{ij}^k}{2}$ with a slight reduction of accuracy (To be checked numerically). 
Because Equations for $S$ and $I$ in Eq.~(\ref{eq_CN-ADI1}) are solved alternatively in time-discretization, we can focus on the solution of $S$ to derive its ADI scheme.  The ADI scheme for $I$ can be derived similarly. 

The equation for $S$ in Eq.~(\ref{eq_CN-ADI1}) can be rewritten as 
\begin{equation}
\displaystyle
(1+\frac{\Delta t}{2} I_{ij}^{k+\frac{1}{2}}  )S_{ij}^{k+1} - \frac{\Delta t}{2} (D_{xx}+D_{yy}) S_{ij}^{k+1} = 
(1- \frac{\Delta t}{2} I_{ij}^{k+\frac{1}{2}}  )S_{ij}^{k} + \frac{\Delta t}{2} (D_{xx}+D_{yy}) S_{ij}^{k}
+ \Delta t f_{ij}^{k+\frac{1}{2}}
\label{eq_CNADI2}
\end{equation}
Based on this equation, we propose the following Peaceman-Rachford \cite{Peaceman:1955} ADI scheme to approximate solution $S_{ij}^{k+1}$ of Eq.~(\ref{eq_CNADI2})
\begin{eqnarray}
\displaystyle\label{eq_cNADI31}
(1+\frac{\Delta t}{4}I_{ij}^{k+\frac{1}{2}}  -  \frac{\Delta t}{2}D_{xx})S_{ij}^{k^*} &=& 
(1-\frac{\Delta t}{4}I_{ij}^{k+\frac{1}{2}} +  \frac{\Delta t}{2}D_{yy})S_{ij}^{k} + \frac{\Delta t}{2} f_{ij}^{k+\frac{1}{2}}\\
(1+\frac{\Delta t}{4}I_{ij}^{k+\frac{1}{2}}  -  \frac{\Delta t}{2}D_{yy})S_{ij}^{k+1} &=& 
(1-\frac{\Delta t}{4}I_{ij}^{k+\frac{1}{2}} +  \frac{\Delta t}{2}D_{xx})S_{ij}^{k^*} + \frac{\Delta t}{2} f_{ij}^{k+\frac{1}{2}}\label{eq_cNADI32}
\end{eqnarray}
In Eqs.~(\ref{eq_cNADI31}) and (\ref{eq_cNADI32}), as 1-D subproblems the resulting linear algebraic matrices in solving $S_{ij}^{k^*}$ and $S_{ij}^{k+1}$ are tridiagonal thus can be solved by Thomas's algorithm with computational cost of $O(N^2)$, which is faster than the $O(N^3)$ SOR. 

To derive this scheme, we plug Eq.~(\ref{eq_cNADI31})$\times (1-\frac{\Delta t}{4}I_{ij}^{k+\frac{1}{2}} +  \frac{\Delta t}{2}D_{xx})$ into Eq.~(\ref{eq_cNADI32})$\times (1+\frac{\Delta t}{4}I_{ij}^{k+\frac{1}{2}}  -  \frac{\Delta t}{2}D_{xx})$ to eliminate the terms containing $S_{ij}^{k^*}$:\\
\begin{eqnarray}\nonumber
& & (1+\frac{\Delta t}{4}I_{ij}^{k+\frac{1}{2}}  -  \frac{\Delta t}{2}D_{xx})(1+\frac{\Delta t}{4}I_{ij}^{k+\frac{1}{2}}  -  \frac{\Delta t}{2}D_{yy})S_{ij}^{k+1} \\ 
&=& 
(1-\frac{\Delta t}{4}I_{ij}^{k+\frac{1}{2}} +  \frac{\Delta t}{2}D_{xx})(1-\frac{\Delta t}{4}I_{ij}^{k+\frac{1}{2}} +  \frac{\Delta t}{2}D_{yy})S_{ij}^{k} + \Delta t  f_{ij}^{k+\frac{1}{2}} \label{eq_CNADI4}
\end{eqnarray}
Expand both sides of Eq.~(\ref{eq_CNADI4}) and rearrange terms by the order of $\Delta t$, we have 
\begin{equation}
(1+\frac{\Delta t}{2}I_{ij}^{k+\frac{1}{2}} -  \frac{\Delta t}{2}D_{xx} - \frac{\Delta t}{2}D_{yy})S_{ij}^{k+1} = 
(1- \frac{\Delta t}{2}I_{ij}^{k+\frac{1}{2}} +  \frac{\Delta t}{2}D_{xx} + \frac{\Delta t}{2}D_{yy})S_{ij}^{k} + \Delta t  f_{ij}^{k+\frac{1}{2}} + O((\Delta t)^3)
\end{equation} 
where the $O((\Delta t)^3)$ term is given as:
\begin{equation}
\left(\frac{(\Delta t)^2}{16}(I_{ij}^{k+\frac{1}{2}})^2-\frac{(\Delta t)^2}{8}I_{ij}^{k+\frac{1}{2}}D_{xx} - \frac{(\Delta t)^2}{8}I_{ij}^{k+\frac{1}{2}}D_{yy} + \frac{(\Delta t)^2}{4}D_{xx}D_{yy}\right)(S_{ij}^k - S_{ij}^{k+1})
\end{equation} 
Note $(S_{ij}^k - S_{ij}^{k+1}) = O(\Delta t) $.\\ 

Similarly, we can drive the ADI scheme for $I$ in Eq.~(\ref{eq_CN-ADI1}), and it can be written as: 

\begin{equation}
\displaystyle
[1 + \frac{\Delta t}{2} (\lambda-S_{ij}^{k})  - \frac{\Delta t}{2} (D_{xx}+D_{yy})]  
 I_{ij}^{k+\frac{1}{2}} = 
[1 - \frac{\Delta t}{2} (\lambda-S_{ij}^{k})  + \frac{\Delta t}{2} (D_{xx}+D_{yy})]  
 I_{ij}^{k-\frac{1}{2}}
+ \Delta t g_{ij}^{k}
\label{eq_CNADI5}
\end{equation}
Following the same procedure of $S$ as above, we can derive the ADI scheme to approximate solution $I_{ij}^{k+\frac{1}{2}}$ of Eq.~(\ref{eq_CNADI5})
\begin{eqnarray}
\displaystyle\label{eq_cNADI61}
(1+\frac{\Delta t}{4}(\lambda-S_{ij}^{k})  -  \frac{\Delta t}{2}D_{xx})I_{ij}^{k^*} &=& 
(1-\frac{\Delta t}{4}(\lambda-S_{ij}^{k})  +  \frac{\Delta t}{2}D_{yy})I_{ij}^{k-\frac{1}{2}} + \frac{\Delta t}{2} g_{ij}^{k}\\
(1+\frac{\Delta t}{4}(\lambda-S_{ij}^{k})  -  \frac{\Delta t}{2}D_{yy}) I_{ij}^{k+\frac{1}{2}} &=& 
(1-\frac{\Delta t}{4}(\lambda-S_{ij}^{k}) +  \frac{\Delta t}{2}D_{xx})I_{ij}^{k^*} + \frac{\Delta t}{2} g_{ij}^{k}\label{eq_cNADI62}
\end{eqnarray}
In Eqs.~(\ref{eq_cNADI61}) and (\ref{eq_cNADI62}), as 1-D subproblems the resulting linear algebraic matrices in solving $I_{ij}^{k^*}$ and $I_{ij}^{k+\frac{1}{2}}$ are tridiagonal thus can be solved by Thomas's algorithm with computational cost of $O(N^2)$, which is faster than the $O(N^3)$ SOR.\\ 
To derive this scheme, we plug Eq.~(\ref{eq_cNADI61})$\times [1-\frac{\Delta t}{4}(\lambda-S_{ij}^{k}) +  \frac{\Delta t}{2}D_{xx}]$ into Eq.~(\ref{eq_cNADI62})$\times [1+\frac{\Delta t}{4}(\lambda-S_{ij}^{k})  -  \frac{\Delta t}{2}D_{xx}]$ to eliminate the terms containing $I_{ij}^{k^*}$:
\begin{eqnarray}\nonumber
& & [1+\frac{\Delta t}{4}(\lambda-S_{ij}^{k})  -  \frac{\Delta t}{2}D_{xx}][1+\frac{\Delta t}{4}(\lambda-S_{ij}^{k})  -  \frac{\Delta t}{2}D_{yy}]I_{ij}^{k+\frac{1}{2}} \\ 
&=& 
[1-\frac{\Delta t}{4}(\lambda-S_{ij}^{k}) +  \frac{\Delta t}{2}D_{xx}][1-\frac{\Delta t}{4}(\lambda-S_{ij}^{k}) +  \frac{\Delta t}{2}D_{yy}]I_{ij}^{k-\frac{1}{2}} + \Delta t  g_{ij}^{k} \label{eq_CNADI7}
\end{eqnarray}
Expand both sides of Eq.~(\ref{eq_CNADI7}) and rearrange terms by the order of $\Delta t$, we have 
\begin{equation}
[1+\frac{\Delta t}{2}(\lambda-S_{ij}^{k}) -  \frac{\Delta t}{2}(D_{xx} + D_{yy})]I_{ij}^{k+\frac{1}{2}} = 
[1- \frac{\Delta t}{2}(\lambda-S_{ij}^{k}) +  \frac{\Delta t}{2}(D_{xx} + D_{yy})]I_{ij}^{k-\frac{1}{2}} + \Delta t  g_{ij}^{k} + O((\Delta t)^3)
\end{equation} 
where the $O((\Delta t)^3)$ term is given as:
\begin{equation}
\left[\frac{(\Delta t)^2}{16}(\lambda-S_{ij}^{k})^2-\frac{(\Delta t)^2}{8}(\lambda-S_{ij}^{k})(D_{xx} + D_{yy}) + \frac{(\Delta t)^2}{4}D_{xx}D_{yy}\right](I_{ij}^{k-\frac{1}{2}} - I_{ij}^{k+\frac{1}{2}})
\end{equation} 
Where $(I_{ij}^{k-\frac{1}{2}} - I_{ij}^{k+\frac{1}{2}}) = O(\Delta t) $.

\subsubsection{Thomas algorithm for a tridiagonal system} 

\noindent
To solve the following tridiagonal system:\\
$\displaystyle{
\begin{pmatrix}
b_1 & c_1 &    &         &          \cr
a_2 & b_2 & c_2 &         &          \cr
   & \ddots & \ddots & \ddots &          \cr
   &            & \ddots & \ddots & c_{n-1} \cr
   &            &        & a_n & b_n \cr \end{pmatrix} =
   \begin{pmatrix}
1 & &    &         &          \cr
l_2 & 1 & &         &          \cr
   & \ddots & \ddots & &          \cr
   &            & \ddots & \ddots & \cr
   &            &        & l_n & 1 \cr\end{pmatrix}
   \begin{pmatrix}
u_1 & c_1 &    &         &          \cr
   & u_2 & c_2 &         &          \cr
   & & \ddots & \ddots &          \cr
   &            & & \ddots & c_{n-1} \cr
   &            &        &   & u_n\cr \end{pmatrix} }$\\
we can use the following procedure.\\

\noindent
\underbar{find $L , U$}\\
\noindent
$b_1 = u_1 \hskip 0.61in ~\Rightarrow~ u_1 = b_1$\\
\noindent
$\!\!\left.\begin{matrix}
a_k = l_k u_{k-1} \hfill & \!\!\Rightarrow~ l_k = a_k/u_{k-1} \hfill \cr
b_k = l_k c_{k-1} + u_k & \!\!\Rightarrow~ u_k = b_k - l_k c_{k-1} \cr\end{matrix}\right\}$ ~for~ $k=2:n$ \\
\vskip 5pt
\noindent
\underbar{solve $Lz=r$} \\
\noindent
$z_1 = r_1$ \\
\noindent
$l_k z_{k-1} + z_k = r_k ~\Rightarrow~
z_k = r_k - l_k z_{k-1}$ ~for~ $k=2:n$ \\
\vskip 5pt
\noindent
\underbar{solve $Uw = z$} \\
\noindent
$u_n w_n = z_n \hskip 0.61in ~\Rightarrow~ w_n = z_n/u_n$ \\
\noindent
$u_k w_k + c_k w_{k+1} = z_k ~\Rightarrow~
w_k = (z_k-c_k w_{k+1})/u_k$ ~for~ $k = n-1\,{:}\,-1\,{:}\,1$ \\
\vskip 1pt
\noindent
This algorithm has operation count = $O(N)$ and memory = $O(N)$ if vectors are used instead of full matrices \\

\newpage
We next provide numerical results from solving the Fisher's model. Test cases are designed to confirm that the designed accuracy and efficiency of the solver has been achieved. For cases with possibly analytical solutions, we shown convergence in accuracy and order by using tabular data. Other results will be visualized using sampled plots in forms of surface, scatter, contour at different time and in animation.

\section{Results}
The numerical simulation are implemented using python on Macbook Pro laptop owned by the students. We will report the machine configuration is CPU time is reported.  
\noindent
\subsection{Convergence test for solving 2-D Fisher's equation with a true solution}

\noindent
To make a test case with true solution, we start with some smooth true values of $S(x,y,t)$ and $I(x,y,t)$ as
\begin{equation}
\displaystyle \left\{\begin{array}{ll} 
S(x,y,t) = \cos(xy) e^t \\
\\
I(x,y,t) = e^x \sin(y) \cos(t) \\
\end{array}\right.
\end{equation}

Plugging into the 2-D Fisher's equation as in Eq.~(\ref{eq_Fisher2d}), the source terms to balance the equation can be calculated as:
\begin{equation}
\displaystyle \left\{\begin{array}{ll} 
f(x,y,t) = \cos(xy) (1+e^x \sin(y) \cos(t)+y^2+x^2) e^t \\
\\
g(x,y,t) = -e^x \sin(y) (\sin(t)+\cos(t) \cos(xy) e^t) \\
\end{array}\right.
\end{equation}

Consider the evolution of the solution from $t=t_0$ to $t=T$ on the rectangular domain $x \in [x_a, x_b]$ and $y \in [y_a, y_b]$, 
we can set initial condition $S(x,y,t_0)$, $I(x,y,t_0)$ and boundary conditions $S(x_a,y,t)$, $I(x_a,y,t)$, $S(x_b,y,t)$, $I(x_b,y,t)$, $S(x,y_a,t)$, $I(x,y_a,t)$, $S(x,y_b,t)$, $I(x,y_b,t)$. \\

%

\noindent
\underline{Part I: Convergence in time discretization}\\

\begin{table}[htp]
\caption{Temporal Convergence (fixed $\Delta h = 0.015625$, variable $\Delta t$ )}
\begin{center}
\begin{tabular}{c|cc|cc|cc|cc}
\hline
& \multicolumn{4}{|c|}{Implicit Euler} & \multicolumn{4}{|c}{Staggered CN} \\\hline
& \multicolumn{2}{|c|}{S} &  \multicolumn{2}{|c|}{I}  & \multicolumn{2}{|c|}{S} &  \multicolumn{2}{|c}{I} \\\hline
$\Delta t$ & Error & Order & Error & Order & Error & Order & Error & Order\\\hline
1/2  & 5.54E-03 & - & 2.56E-02 & - & 3.86E-03 & - & 1.28E-02 & - \\
1/4  & 3.22E-03 & 0.78 & 1.43E-02 & 0.84 & 9.48E-04 & 2.02 & 3.25E-03 & 1.98 \\
1/8  & 1.72E-03 & 0.91 & 7.63E-03 & 0.91 & 2.38E-04 & 1.99 & 8.13E-04 & 2.00 \\
1/16 & 8.81E-04 & 0.96 & 3.94E-03 & 0.95 & 5.93E-05 & 2.01 & 2.03E-04 & 2.00 \\
1/32 & 4.46E-04 & 0.98 & 2.00E-03 & 0.98 & 1.45E-05 & 2.03 & 5.02E-05 & 2.01 \\
1/64 & 2.24E-04 & 0.99 & 1.01E-03 & 0.99 & 3.27E-06 & 2.15 & 1.21E-05 & 2.05\\\hline
\end{tabular}
\end{center}
\label{default}
\end{table}%

\begin{table}[htp]
\caption{Temporal Convergence (fixed $\Delta h = 0.015625$, variable $\Delta t$, SOR weight $\omega = \frac{2}{(1 + \sqrt{1 - cos^2(\pi h)}}$ )}

{\scriptsize
\begin{center}
\begin{tabular}{c|cc|cc|c|cc|cc|c|cc|cc|c}
\hline
& \multicolumn{5}{|c|}{Full matrix} & \multicolumn{5}{|c}{SOR Iterative} &\multicolumn{5}{|c}{ADI} \\\hline
& \multicolumn{2}{|c|}{S} &  \multicolumn{3}{|c|}{I}  & \multicolumn{2}{|c|}{S} &  \multicolumn{3}{|c}{I} & \multicolumn{2}{|c|}{S} &  \multicolumn{3}{|c}{I}\\\hline
$\Delta t$ & err. & ord. & err. & ord. & time & err. & ord. & err. & ord. & time & err. & ord. & err. & ord. & time\\\hline
1/2  &  3.9e-3 &       - & 1.3e-2 & - & 1.6 & 3.7e-3 & - & 1.3e-2 & - & 6.3 & 7.0e-2 & - & 1.0e-2 & - & 0.08 \\
1/4  &  9.5e-4 & 2.0 & 3.3e-3 & 2.0 & 3.0 & 9.2e-4 & 2.0 & 3.3e-3 & 2.0 & 11.2 & 2.3e-2 & 1.6 & 3.0e-3 & 1.7 & 0.15 \\
1/8  &  2.4e-4 & 2.0 & 8.1e-4 & 2.0 & 6.0 & 2.3e-4 & 2.0 & 8.4e-4 & 2.0 & 21.3 & 6.1e-3 & 1.9 & 8.4e-4 & 1.9 & 0.27 \\
1/16 & 5.9e-5 & 2.0 & 2.0e-4 & 2.0 & 11.5 & 5.7e-5 & 2.0 & 2.1e-4 & 2.0 & 41.6 & 1.5e-3 & 2.0 & 2.2e-4 & 2.0 & 0.52 \\
1/32 & 1.5e-5 & 2.0 & 5.0e-5 & 2.0 & 22.1 & 1.4e-5 & 2.0 & 5.2e-5 & 2.0 & 83.2 & 3.9e-4 & 2.0 & 5.4e-5 & 2.0 & 1.01 \\
1/64 & 3.3e-6 & 2.1 & 1.2e-5 & 2.1 & 43.3 & 3.2e-6 & 2.1 & 1.2e-5 & 2.0 & 165.6 & 9.7e-5 & 2.0 & 1.3e-5 & 2.1 & 2.01 \\\hline
\end{tabular}
\end{center}
}
\label{default}
\end{table}%

\noindent
\underline{Part I: Convergence in space discretization}\\

\begin{table}[htp]
\caption{Spacial Convergence (fixed $\Delta t = 0.0001$, variable $\Delta h$ )}
\begin{center}
\begin{tabular}{c|cc|cc|cc|cc}
\hline
& \multicolumn{4}{|c|}{Implicit Euler} & \multicolumn{4}{|c}{Staggered CN} \\\hline
& \multicolumn{2}{|c|}{S} &  \multicolumn{2}{|c|}{I}  & \multicolumn{2}{|c|}{S} &  \multicolumn{2}{|c}{I} \\\hline
$\Delta t$ & Error & Order & Error & Order & Error & Order & Error & Order\\\hline
1/2 & &&&& 1.58E-04 & - & 1.45E-03 & - \\
1/4  & 1.23E-04 & - & 2.89E-04 & - & 		1.24E-04 & 0.35 & 2.97E-04 & 2.29 \\
1/8  & 3.15E-05 & 1.96 & 5.89E-05 & 2.30 & 3.26E-05 & 1.93 & 6.62E-05 & 2.17 \\
1/16 & 7.03E-06 & 2.17 & 8.74E-06 & 2.75 & 7.93E-06 & 2.04 & 1.56E-05 & 2.09 \\
1/32 & 1.47E-06 & 2.25 & 2.85E-06 & 1.62 & 1.94E-06 & 2.03 & 3.78E-06 & 2.04\\
1/64 &&&&& 4.78E-07 & 2.02 & 9.30E-07 & 2.02 \\ \hline

\end{tabular}
\end{center}
\label{default}
\end{table}%

\begin{table}[htp]
\caption{Spacial Convergence (fixed $\Delta t = 0.0001$, variable $\Delta h$)}
{\scriptsize
\begin{center}
\begin{tabular}{c|cc|cc|c|cc|cc|c|cc|cc|c}
\hline
& \multicolumn{5}{|c|}{Full matrix} & \multicolumn{5}{|c}{Gauss-Seidel Iterative} &\multicolumn{5}{|c}{ADI} \\\hline
& \multicolumn{2}{|c|}{S} &  \multicolumn{3}{|c|}{I}  & \multicolumn{2}{|c|}{S} &  \multicolumn{3}{|c}{I} & \multicolumn{2}{|c|}{S} &  \multicolumn{3}{|c}{I}\\\hline
$\Delta h$ & err. & ord. & err. & ord. & time & err. & ord. & err. & ord. & time & err. & ord. & err. & ord. & time\\\hline
1/2 & 1.6e-4 & - & 1.5e-3 & - & 0.4 & 5.3e-5 & - & 4.9e-4 & - & 0.3 & 5.2e-7 & - & 4.9e-4 & - & 0.4\\
1/4  &  1.2e-4 & 0.4 & 3.0e-4 & 2.3 & 1.5 & 7.4e-5 & -0.5 & 1.8e-4 & 1.5 & 1.8 & 7.4e-5 & -0.5 & 1.8e-4 & 1.5 & 1.3 \\
1/8  &  3.3e-5 & 1.9 & 6.6e-5 & 2.2 & 14.9 & 2.5e-5 & 1.6 & 5.1e-5 & 1.8 & 9.2 & 2.5e-5 & 1.6 & 5.1e-5 & 1.8 & 4.8 \\
1/16  &  7.9e-6 & 2.0 & 1.6e-5 & 2.1 & 56.2 & 7.0e-6 & 1.9 & 1.4e-5 & 1.9 & 46.1 & 7.0e-6 & 1.9 & 1.4e-5 & 1.9 & 19.4 \\
1/32 & 1.9e-6 & 2.0 & 3.8e-6 & 2.0 & 329.4 & 1.8e-6 & 1.9 & 3.5e-6 & 2.0 & 250.1 & 1.8e-6 & 1.9 & 3.5e-6 & 1.9 & 78.2 \\
1/64 & 4.8e-7 & 2.0 & 9.3e-7 & 2.0 & 6655.8 & 4.6e-7 & 2.0 & 9.0e-7 & 2.0 & 1731.9 & 4.6e-7 & 2.0 & 9.0e-7 & 2.0 & 324.0 \\
\hline

\end{tabular}
\end{center}
}
\label{default}
\end{table}%

Time calculated by Apple M2 Pro.\\\\

\begin{figure}[htb!]
\includegraphics[width=0.45\textwidth,angle=0]{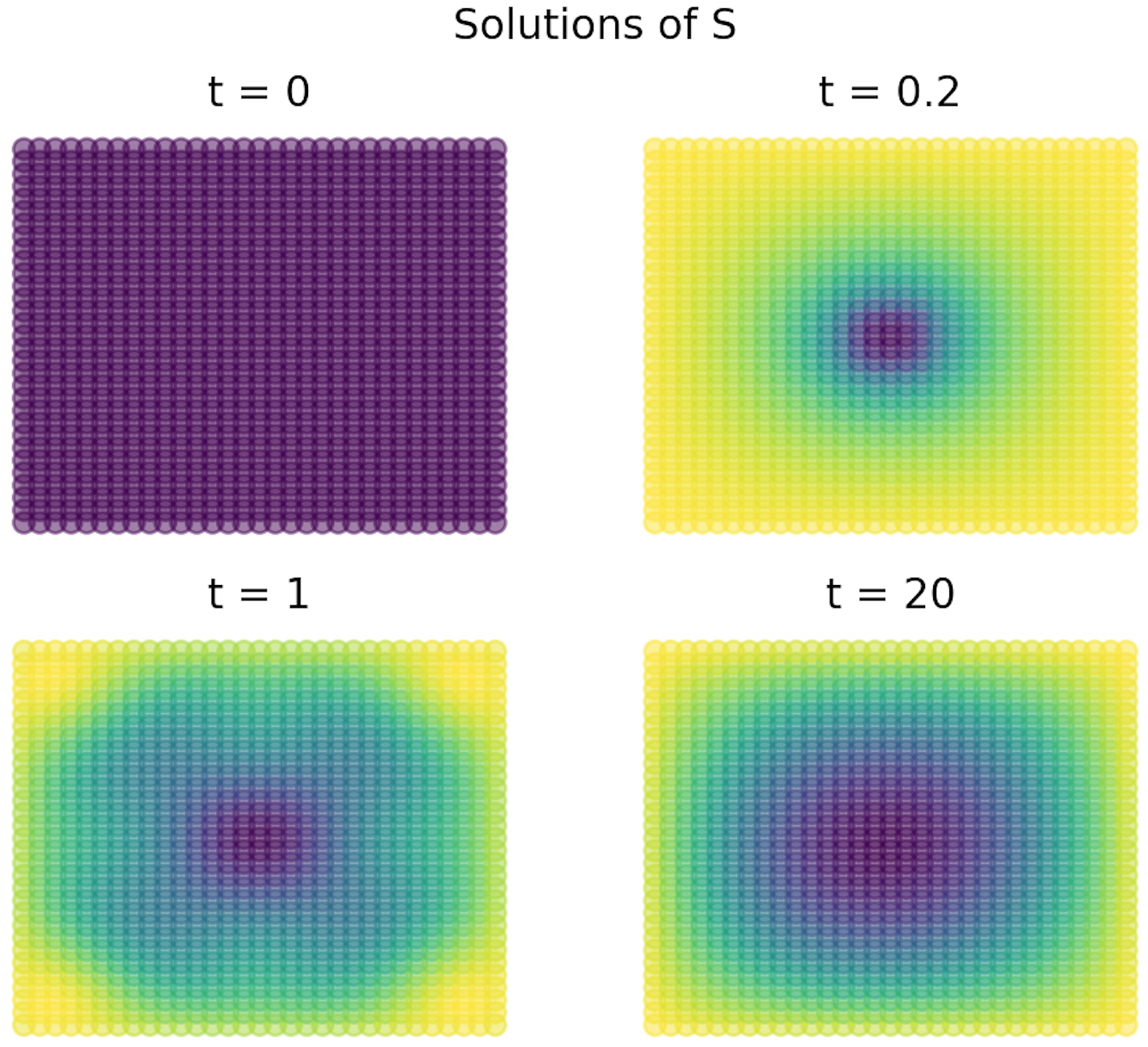} \hskip 0.4in
\includegraphics[width=0.45\textwidth,angle=0]{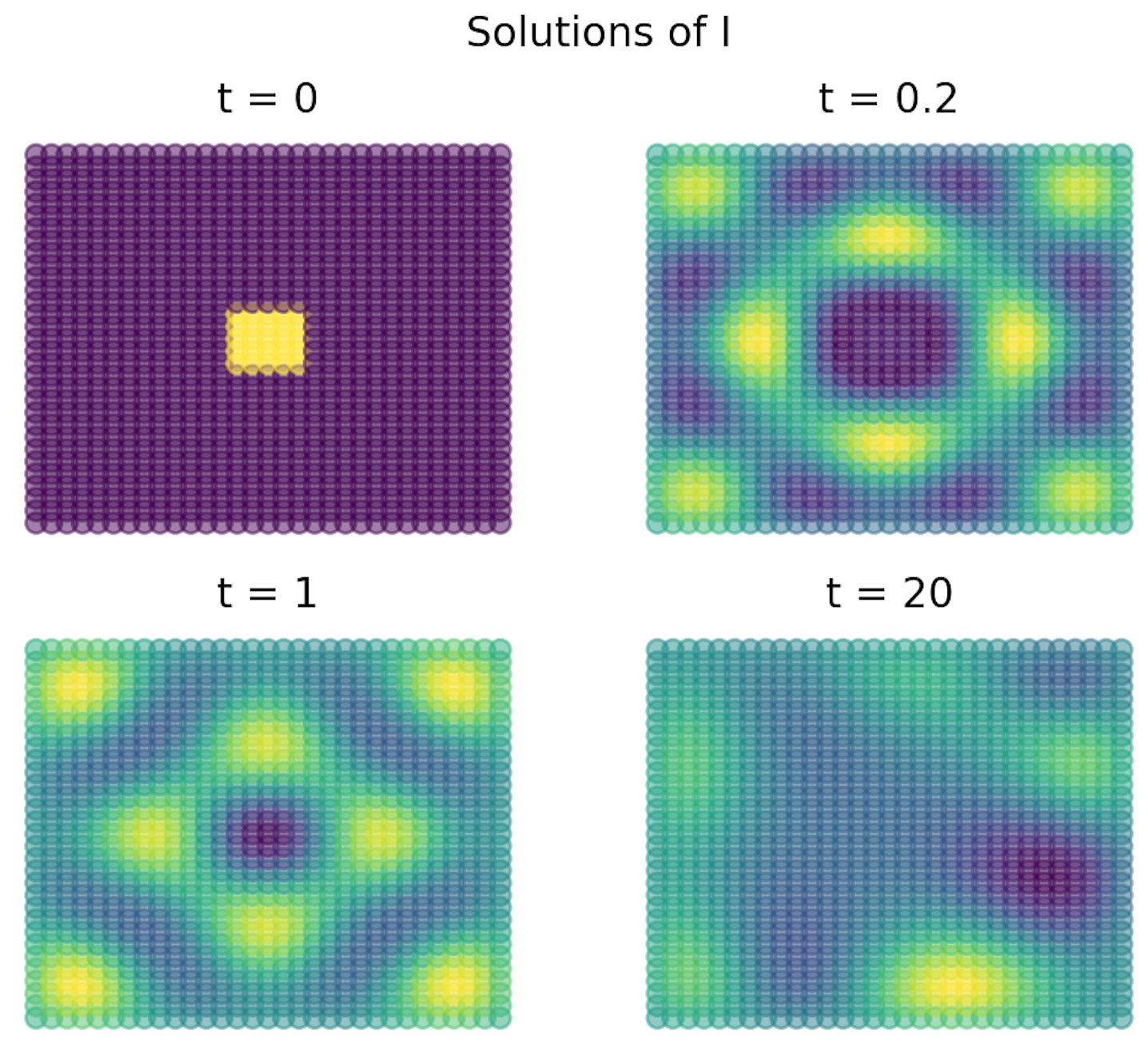}
\begin{center}
(a) \hskip 3.0in (b)\\
\end{center}
\vskip -10pt
\caption{Numerical solutions in space at different time  (a) the susceptible group (S) (b) the infectious group (I) with initially non-zero values at the center of the region.}
\label{fig_1}
\end{figure}

We illustrate a very preliminary result in Fig.~(\ref{fig_1}) for solving the Fisher's equation with initial condition for $S$ as a constant
and 
a small centered portion of support for $I$ indicating the origin of Covid-19 locally. The boundary conditions are constants for simplification. The numerical simulation from this simple can already show the spreading pattern the disease.  
This Fisher's model can be easily converted to the SI model by setting $S$ and $I$ only time-dependent without diffusion. Essentially, the SI model shows the logistical growth/decay (e.g. sigmoid curves) of the population/density, while the Fisher's model involves the diffusion for spacial spreading of the disease. We will focus how to numerically solve the Fisher's model subject to various initial and boundary conditions. \\

\subsubsection{Boundary Conditions}

\noindent
\underline{Part II: Apply the models using real world Covid-19 Data}  \\
There are abundant and well-maintained Covid-19 data available online, updated daily in scales from counties to countries. We use out model disease spreading in regions as different levels. We will design source terms like $f(t, x,y)$ and $g(t,x,y)$ and boundary and initial conditions according to the physical environment. Known data will be used to fit the parameters such as reaction rate and diffusion rate. Predicted data will be available once the model is ready. Inference will be drawn based on the simulation and available data.

\section{Conclusion}

\section*{Acknowledgements}
Talha and Geng's research is partially supported by the National Science Foundation (NSF) under grant DMS-2110922.
Zhao’s research is partially supported by the NSF under grants DMS-2110914.
Xu's research is partially supported by SMU Hamilton's Scholar and URA programs. 

\bibliographystyle{acm}
\bibliography{ref_clock}

\begin{thebibliography}{1}

\bibitem{Fisher:1937}
{\sc Fisher, R.~A.}
\newblock The wave of advance of advantageous genes.
\newblock {\em Annals of Eugenics 7\/} (1937), 355--369.

\bibitem{Habbal:2014}
{\sc Habbal, A., Barelli, H., and Malandain, G.}
\newblock Assessing the ability of the 2d fisher–kpp equation to model
  cell-sheet wound closure.
\newblock {\em Mathematical Biosciences 252\/} (2014), 45--59.

\bibitem{Kenkre:2024}
{\sc Kenkre, V.}
\newblock Results from variants of the fisher equation in the study of
  epidemics and bacteria.
\newblock {\em Physica A: Statistical Mechanics and its Applications 342}, 1
  (2004), 242--248.
\newblock Proceedings of the VIII Latin American Workshop on Nonlinear
  Phenomena.

\bibitem{Li:2009}
{\sc Li, J., and Zou, X.}
\newblock Modeling spatial spread of infectious diseases with a fixed latent
  period in a spatially continuous domain.
\newblock {\em Bulletin of Mathematical Biology 71}, 8 (2009), 2048--2079.

\bibitem{Lou:2023}
{\sc Lou, Y., and Salako, R.~B.}
\newblock Mathematical analysis of the dynamics of some reaction-diffusion
  models for infectious diseases.
\newblock {\em Journal of Differential Equations 370\/} (2023), 424--469.

\bibitem{Peaceman:1955}
{\sc Peaceman, D.~W., and Rachford, Jr., H.~H.}
\newblock The numerical solution of parabolic and elliptic differential
  equations.
\newblock {\em Journal of the Society for Industrial and Applied Mathematics
  3}, 1 (1955), 28--41.

\bibitem{Vaziry:2022}
{\sc Vaziry, A., Kolokolnikov, T., and Kevrekidis, P.~G.}
\newblock Modelling of spatial infection spread through heterogeneous
  population: from lattice to partial differential equation models.
\newblock {\em Royal Society Open Science 9}, 10 (2022), 220064.

\end{thebibliography}
\end{document}